\documentclass{jfm}

\usepackage{graphicx}
\usepackage{color}
\usepackage{amsmath}
\usepackage{float}
\usepackage{tabularx}

\begin{document}

\shorttitle{Modelling of an actuated elastic swimmer} 
\shortauthor{Pi\~neirua, Thiria and Godoy-Diana} 

\title{Modelling of an actuated elastic swimmer}

\author
 {
 M. Pi\~neirua\aff{1}$^{,}$\aff{2}$^{,\corresp{\email{miguel.pineirua@univ-tours.fr }}}$,
  B. Thiria\aff{1}$^{,\corresp{\email{bthiria@pmmh.espci.fr}}}$
  \and 
  R. Godoy-Diana\aff{1}$^{,\corresp{\email{ramiro@pmmh.espci.fr}}}$
  }

\affiliation
{
\aff{1}
Laboratoire de Physique et M\'ecanique des Milieux H\'et\'erog\`enes (PMMH), CNRS UMR 7636, ESPCI Paris---PSL Research University, Sorbonne Universit\'es---Universit\'e Pierre et Marie Curie---Paris 6, Universit\'e Paris Diderot---Paris 7, 10 rue Vauquelin, 75005 Paris, France.
\aff{2}
Institut de Recherche sur la Biologie de l'Insecte (IRBI), CNRS UMR 7261, UFR Sciences et Techniques, Universit\'e Fran\c cois Rabelais, 37200 Tours, France
}

\maketitle

\begin{abstract}
We study the force production dynamics of undulating elastic plates as a model for fish-like inertial swimmers. Using a beam model coupled with Lighthill's large-amplitude elongated-body theory, we explore different localised actuations at one extremity of the plate (heaving, pitching, and a combination of both) in order to quantify the reactive and resistive contributions to the thrust. The latter has the form of a quadratic drag in large Reynolds number swimmers and has recently been pointed out as a crucial element in the thrust force balance. We validate the output of a weakly nonlinear solution to the fluid--structure model using thrust force measurements from an experiment with flexible plates subjected to the three different actuation types. The model is subsequently used in a self-propelled configuration ---with a skin friction model that balances thrust to produce a constant cruising speed--- to map the reactive versus resistive thrust production in a parameter space defined by the aspect ratio and the actuation frequency. We show that this balance is modified as the frequency of excitation changes and the response of the elastic plate shifts between different resonant modes, the pure heaving case being the most sensitive to the modal response with drastic changes in the reactive/resistive contribution ratio along the frequency axis. We analyse also the role of the phase lag between the heaving and pitching components in the case of combined actuation, showing in particular a non-trivial effect on the propulsive efficiency.
\end{abstract}

\section{Introduction}

Fish-like swimmers use muscle action to produce the undulations that give rise to a propulsive force \citep{Blake:2004}.  However, this active driving of the kinematics does not give the full picture of the animal swimming problem, since the elastic properties of fins introduce a passive component to the balance of internal forces during the flapping cycle. Flexural stiffness and body geometry interact in non-trivial ways to determine swimming performance \cite[see e.g.][]{Lucas:2014,Tytell:2014,vanWeerden:2014,Feilich:2015} and are crucial for the description of the role of passive appendages or artificial swimmers. Elastic plates \cite[][]{Alben:2012,Dewey:2013,Raspa:2014,Paraz:2014,Quinn:2014,Quinn:2015,Fernandez-Prats:2015,Paraz:2016} or rods \citep{Ramananarivo:2013,Ramananarivo:2014b} with localised actuation have been shown recently to be a versatile model to study the dynamics of undulatory swimming. In these cases, the deformation kinematics is not actively enforced, but is fully given by the body elasticity. The dynamics of such elastic swimmers constitutes a strongly coupled fluid--structure interaction problem. The net force on the swimmer is of course given at each instant by the integration of the pressure field and viscous stress over the body surface. For large Reynolds number ($\Rey=UL/\nu$ ) flows $\Rey\gg 1$ inertia dominates and the effects of viscosity are confined to a thin boundary layer around the swimmer. The dynamics outside the boundary layer is then described by the Euler equations for a perfect fluid.  When $\Rey\ll 1$, it is the viscous term alone that balances the pressure gradient in the Navier--Stokes equation. This limit, known as Stokes flow, describes for instance the propulsion of microscopic organisms using cilia or flagella \cite[see e.g.][]{Lauga:2009}. Aquatic locomotion spans the whole Reynolds number spectrum, from one limit to the other, bringing especially rich problems in the intermediate cases where conventional analytical methods fail \citep{Childress:1981}. The comprehension of this intermediate range of Reynolds numbers requires the correct modelling of the vortex dynamics detaching from the swimmer, and considerable efforts in this sense have been widely documented in the literature \cite[see e.g.][and references therein]{Michelin:2008,Alben:2009,Sheng:2012}. 

\cite{Lighthill:1960} and \cite{Wu:1961} established the theoretical foundation to understand reactive force production in a perfect fluid in, respectively, the slender-body limit and the two-dimensional limit. Considering the case of slender bodies has been proven useful to understand the mechanics of undulatory swimmers. The framework of large-amplitude elongated bodies in a perfect fluid of \cite{Lighthill:1971} is in this matter the established theory \cite[see also][]{Candelier:2011}. The aforementioned balance of forces at a cross-section of a slender body needs nonetheless the inclusion of a resistive contribution due to the lateral drag to be complete. Although this resistive \emph{quadratic drag} was already identified by \cite{Taylor:1952} in his analysis of the swimming of long and narrow animals, it is only recently that it has been incorporated alongside the reactive forces as a crucial element of several fluid--structure problems in the inertial regime such as the dynamics of flexible flapping wings \cite[][]{Ramananarivo:2011}, flapping flags \cite[][]{Eloy:2012,Singh:2012} and self-propelled swimmers \cite[][]{Ramananarivo:2013,Eloy:2013,Porez:2014}. Physically, this quadratic drag is associated to the dynamic stalls at each swimming cycle that result from the large transversal local velocities and the finite geometry of the swimmer, as can be observed for instance in numerical simulations of flow around swimming fish \cite[see e.g.][]{Borazjani:2010,vanRees:2013,vanRees:2015,Li:2016}.  It should be noted that the nonlinear character of this resistive term sets it apart from the viscous dissipation that has been included in other models of similar problems \cite[][]{Argentina:2005,Gazzola:2015}. Concerning artificial elastic swimmers with localised actuation, \citet*{Ramananarivo:2014a} showed that the resistive term is determinant for the establishment of the propagative wave that mimics the kinematics of animal swimmers, and \cite{Paraz:2016} have incorporated it in a successful model predicting thrust generation of a heaving foil. On the other hand, \citet*{Pineirua:2015} have quantified the role of the resistive term in the thrust production balance for two typical body kinematics found in nature, anguilliform and carangiform, using a flat plate of different aspect ratios as a model. It is remarkable that, depending on the kinematics and the geometry of the swimmer, the resistive term can outperform the reactive mechanisms as main source of thrust. 

The goal of the present paper is to examine the equilibrium between the reactive and resistive thrust production for the case of locally actuated model elastic swimmers. For this purpose we will use both a complete analytical model and an experimental set-up where a rectangular elastic plate is actuated on one of its ends using small-amplitude pitching and/or heaving oscillations. 

\section{Modelling of an actuated elastic plate in self-propulsion}\label{sec:model}

\begin{figure}
\centering
\begin{tabular}{c@{\hspace{1cm}}c}
\includegraphics[height=0.25\linewidth]{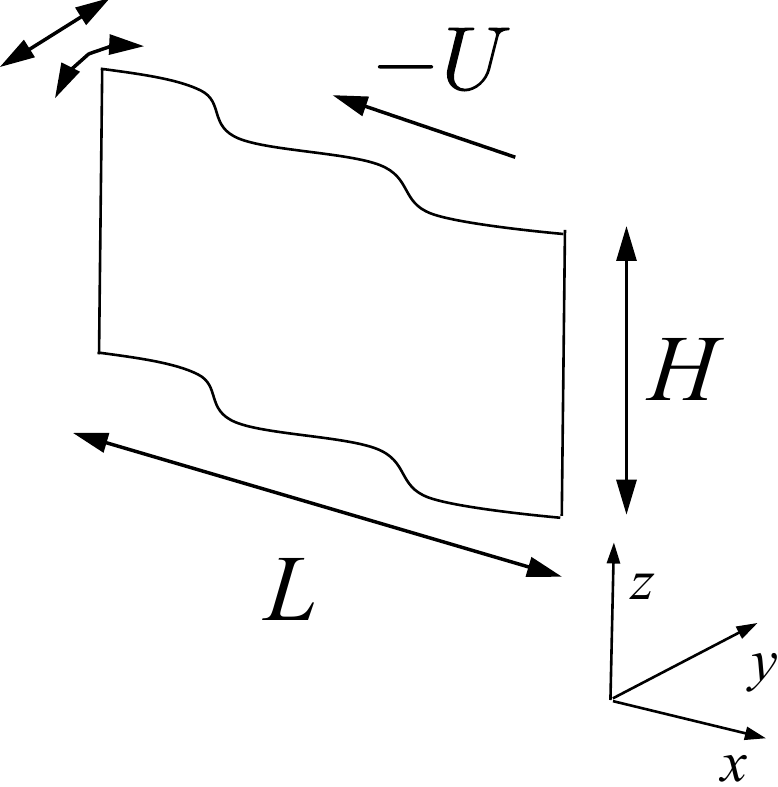}&
\raisebox{.5\height}{\includegraphics[height=0.13\linewidth]{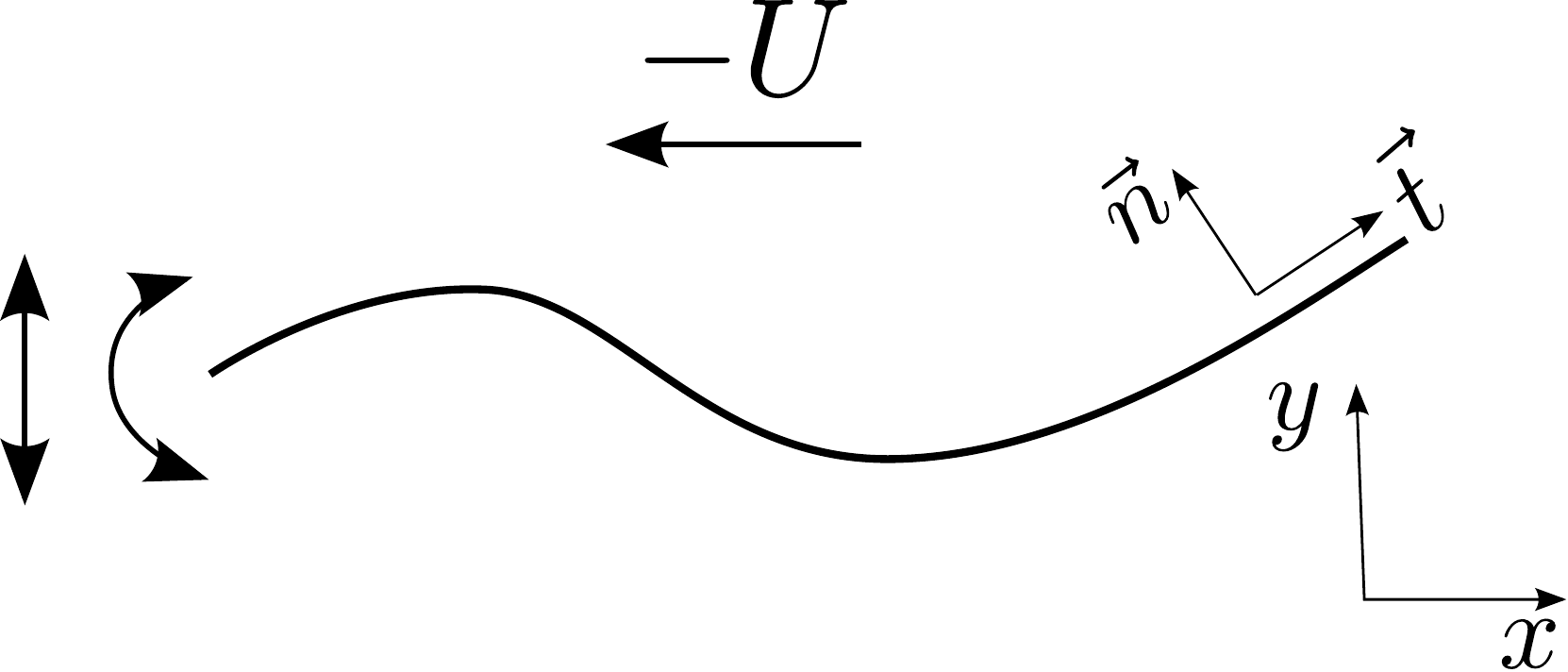}}
\end{tabular}
\caption{Schematic diagram of the elastic plate modelled.}\label{fig:schema}
\end{figure}

We consider a flexible plate of length $L$, span $H$ and negligible thickness, actuated at one of its ends. The geometry of the plate is completely determined by its local orientation with respect to the horizontal axis, $\theta(S,T)$ (see figure \ref{fig:schema}), where $S$ is the curvilinear coordinate along the plate and $T$ is time.  Based on the Euler-Bernoulli beam equation,  the conservation of momentum leads to: 

 \begin{equation}
 \mu\frac{\partial^2 \mathbf{X}}{\partial T^2}=\frac{\partial}{\partial S}\left(F_{\tau}\boldsymbol{\tau}-B\frac{\partial ^2\theta}{\partial S^2}\mathbf{n}\right)-\mathbf{P} ,
 \label{eq_plaque}
\end{equation}

\noindent where $\mu$ is the surface density of the plate, $F_t$ is an internal tension ensuring the inextensibility condition, $B=EI$ is the bending rigidity  and $\mathbf{P}=\mathbf{P}_{am}+\mathbf{P}_{d}$ is the force resulting from the fluid pressure field.  In equation \ref{eq_plaque}, $\mathbf{n}$ and $\boldsymbol{\tau}$ are the unit normal and tangent vectors to the plate (see Figure \ref{fig:schema}).  In the present model  we have neglected any internal viscoelastic dissipation in the beam, considering that damping will be dominated by the external resistive term due to the lateral fluid dynamic drag $\mathbf{P}_{d}$. In addition to this dissipation term, the fluid model is written considering the reactive force due to transverse motions $\mathbf{P}_{am}$ (the $am$ subscript meaning \emph{added mass}).

In the slender-body approximation the fluid forces can be written in terms of the local velocities of each cross-section $U_n$ and $U_t$, related to the swimming velocity $U$ by $U_n\mathbf{n}+U_t\boldsymbol{\tau}=\partial\mathbf{X}/\partial t-U\mathbf{e}_x$. The resistive term is characterised by a lateral drag coefficient $C_d$ and reads \citep{Taylor:1952,Eloy:2012,Singh:2012,Ramananarivo:2013}:

\begin{equation}
\mathbf{P}_{d}=-\frac 1 2 \rho C_d|U_n|U_n\mathbf{n}, 
\label{P_d}
\end{equation}
where $\rho$ is the fluid density. For the lateral Reynolds numbers of the inertial swimmers of interest here, the drag coefficient can be reasonably considered to maintain a constant value of $C_d\sim 2$ \cite[][]{Eloy:2013,Pineirua:2015}.

The reactive term is given by the large-amplitude elongated-body theory of \cite{Lighthill:1971}:

\begin{equation}
\mathbf{P}_{am}=\mathcal{M}(H)\left[\frac{\partial}{\partial S}\left(U_nU_t\mathbf{n}-\frac 1 2 U_n^2\boldsymbol{\tau}\right)-\frac{\partial U_n}{\partial t}\mathbf{n}\right]
\label{P_am}
\end{equation}

\noindent where $\mathcal{M}(H)$ is the added mass due to the fluid, which is a function of the span of the plate \citep{Pineirua:2015}.  For slender-body swimmers, $H/L<0.4$, the added mass can be estimated as $\mathcal{M}(H)=\frac{\pi}{4}\rho H$.

\subsection{Boundary conditions, heave, pitch and pitch--heave}

The boundary conditions are determined according to the actuation applied at the edge of the plate.  In the present study we consider three different types of actuation: pitch, heave and a combination of both.  In each of the three cases, the boundary conditions at the leading edge ($S=0$) are considered as :

\begin{subequations}\label{border_cond}
\begin{align}
 \mathbf{X}(0,T)&=\begin{bmatrix}
 0\\
 A_0\sin (\omega T)
 \end{bmatrix},\; \theta(0,T)=0 \;\;\;\text{for heave},\label{BCheave}\\
 \mathbf{X}(0,T)&=\begin{bmatrix}
 0\\
 0
 \end{bmatrix},\; \theta(0,T)=\psi_0\sin (\omega T) \;\;\;\text{for pitch, and}\label{BCpitch}\\
  \mathbf{X}(0,T)&=\begin{bmatrix}
 0\\
 A_0\sin (\omega T)
 \end{bmatrix},\; \theta(0,T)=\psi_0\sin (\omega T+\phi) \;\;\;\text{for pitch-heave},\label{BCboth}
 \end{align}\end{subequations}
 where $\omega$ is the forcing frequency and $\phi$ is the phase lag between the two components in the pitch and heave case.
 
The boundary conditions at the trailing edge ($S=L$) are independent of the imposed actuation and are kept as torque free and force free, thus, 
\begin{equation}
\frac{\partial\theta}{\partial S}(L,T)=\frac{\partial^2\theta}{\partial S^2}(L,T)=0.
\end{equation}

\subsection{Non-dimensional equations}
\noindent Using $L$ and $1/\omega$,  as characteristic length and time, respectively, equations \eqref{eq_plaque} -- \eqref{P_am} can be re-written in  non-dimensional form as:

\begin{align}
 \frac{\partial^2 \mathbf{x}}{\partial t^2}&=\frac{\partial f_{\tau}}{\partial s}\boldsymbol{\tau}-\frac{1}{\omega^{*2}}\frac{\partial ^3\theta}{\partial s^3}\mathbf{n}-\mathbf{p}, \label{eq_plaque_adim} \\
 \mathbf{p}_{d}&=-\frac 1 2 C_dM^* |U^*_n|U^*_n\,\mathbf{n}, \label{P_d_adim}\\
\mathbf{p}_{am}&=\frac{\pi}{4}H^*M^*\left[\frac{\partial}{\partial s}\left(U^*_nU^*_t\mathbf{n}-\frac 1 2 U^{*2}_n\boldsymbol{\tau}\right)-\frac{\partial U^*_n}{\partial t}\mathbf{n}\right]\label{P_am_adim}.
\end{align}

The problem is characterised by the following non-dimensional parameters:
\begin{align}
&\text {- the non-dimensional frequency}\quad \omega^*=L^2\omega\sqrt{\frac{\mu}{B}},\\ 
&\text {- the fluid-solid inertial ratio}\quad M^*=\frac{\rho L}{\mu},\\ 
&\text {- the non-dimensional plate span}\quad H^*=\frac{H}{L},\\ 
&\text {- the non-dimensional swimming speed}\quad U^*=\frac{U}{L\omega}.
\end{align}

\subsection{Weakly nonlinear approximation and numerical method}
\begin{figure}
\centering
\includegraphics[height=0.25\linewidth]{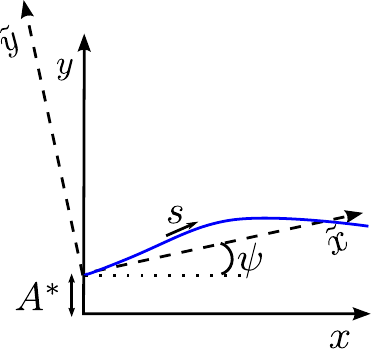}
\caption{Schematic diagram of the reference frame rotation and translation.}\label{fig:axes}
\end{figure}

Following \cite{Eloy:2012}, equation (\ref{eq_plaque_adim}) is projected onto the $x$ and $y$ directions in order to obtain two equations for $x(s,t)$ and $y(s,t)$ respectively.  The horizontal projection is used to eliminate the tension term $f_\tau$ from the $y$ projection.  Finally, $x$ and its derivatives are eliminated using the inextensibility condition.  Keeping terms up to $O(y^3)$ one obtains a weakly nonlinear equation for $y(s,t)$ :

\begin{equation}
L(y)+p_\text{m}(y)+\frac{1}{\omega^{*2}}f_\text{B}(y)+M^* f_{\text{d}}(y)+\frac{\pi}{4}M^* H^* f_{am}(y)=0,
\label{ydest}
\end{equation}
where 
\begin{align}
&L(y)=\ddot{y}+\frac{1}{\omega^{*2}}y^{(4)}+M^*H^*(U^{*2}y''+2U^*\dot{y}'+\ddot{y}),\label{eq:linear}\\
&f_\text{m}(y)=y'\int_0^s(\dot{y}'^2+y'\ddot{y}')ds-y''\int_{s'}^1\int_0^s(\dot{y}'^2+y'\ddot{y}')dsds',\label{eq:nl1}\\
&f_\text{B}(y)=4y'y''y'''+y'^2y^{(4)}+y''^3,\\
&f_{d}(y)=\frac{1}{2}C_d|U^*y'+\dot{y}|(U^*y'+\dot{y}),\\
&f_{am}(y)=-\frac{1}{2}U^{*2}y''y'^2+U^*\dot{y}'y'^2-3U^*y''y'\dot{y}-2\dot{y}'y'\dot{y}-\frac{1}{2}y''\dot{y}^2+y'\int_0^s(\dot{y}'^2+y'\ddot{y}')ds\nonumber \\&+2(U^*y''+\dot{y}')\int_0^s\dot{y}'y'ds-y''\int_s^1y'(U^{*2}y''+2U^*\dot{y}'+\ddot{y})ds.\label{eq:nl5}
\end{align}
Equation~\eqref{eq:linear} corresponds to the linearised dynamics while Equations~\eqref{eq:nl1}--\eqref{eq:nl5} correspond to nonlinearities related to inertia, stiffness, resistive and reactive flow effects, respectively. 

A Galerkin decomposition is used to solve Equation~\eqref{ydest} \cite[see also][]{Pineirua:2015b}: the vertical displacement $\tilde{y}$ on a rotating and translating reference frame (see figure \ref{fig:axes}) is expanded as a superposition of clamped--free beam eigenmodes $\phi_p(s)$,
\begin{equation}
 \tilde{y}(s,t)=\sum_{p=1}^\infty X_p(t)\phi_p(s).
\end{equation}

Translation and rotation of the reference frame are imposed in order to satisfy the leading-edge boundary conditions.  In the small pitch angle approximation, the vertical displacement of the plate can then be written as :

 \begin{equation}
 y(s,t)=\tilde{y}(s,t) + A^*(t)+\tilde{x}(s,t)\psi(t).
 \label{y_tot}
\end{equation}
Next, equation (\ref{y_tot}) is injected into (\ref{ydest}) and then projected on the same set of eigenmodes $\phi_p(s)$. After truncation to $N$ linear modes, the resulting coupled system of equations is integrated numerically using a semi-implicit step-adaptive fourth-order Runge-Kutta method.

\subsection{Self-propelled configuration: skin friction}

Up to now we have described a model in the frame of reference of the swimmer, a configuration that allows for the comparison of thrust production with our experiment where a swimmer is tethered to a force sensor. In a free swimming configuration, however, an additional term enters the force balance in the swimming direction: the skin friction. The equation of motion along the swimming direction can therefore be written for a swimmer of total mass $\mu H L$ as:

\begin{equation}
( \mathbf{p}_{am}+ \mathbf{p}_{d}+\mathbf{p}_{\nu})\cdot \mathbf{e}_x = \frac{\partial U^*}{\partial t} \;, 
\end{equation}

\noindent where $\mathbf{p}_{\nu}$ is the skin friction modified for a flapping plate \cite[see][]{Ehrenstein:2013,Eloy:2013}: 

\begin{equation}
\mathbf{p}_{\nu}\cdot \mathbf{e}_x=\frac{4}{3}\frac{M^*}{\sqrt{Re_ls}}U^{*2}_t e^{-4|U_n/U_t|}-1.4\frac{M^*}{\sqrt{Re_w}}U^*_t |U^*_n|
\label{skin_friction_ehrenstein}
\end{equation}

\noindent where $Re_l=\rho L U_t/\eta$ and $Re_w=\rho HU_n/\eta$ are, respectively, the Reynolds numbers based on the  longitudinal and lateral displacements.  The first term in equation \ref{skin_friction_ehrenstein} when $U_n=0$, corresponds to the friction force experienced by a flat plate with no lateral movement.  The second term takes into account the effect of boundary the layer compression due to the lateral displacement of the plate. 

\begin{figure}
\centering
\includegraphics[width=0.65\textwidth]{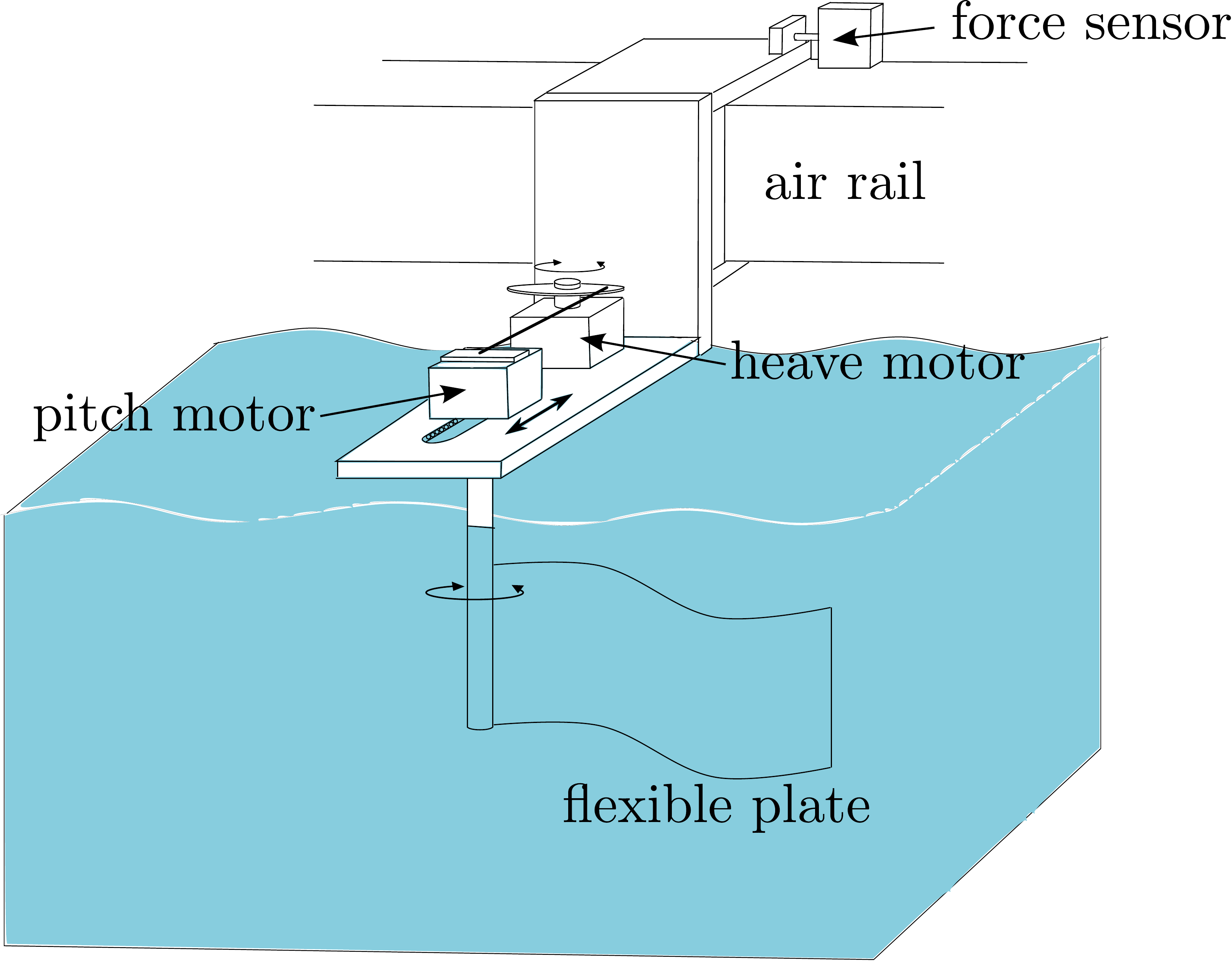}
\caption{Schematic diagram of the experimental setup.}
\label{Fig_setup}
\end{figure}

\begin{figure}
\centering
\includegraphics[width=0.95\textwidth]{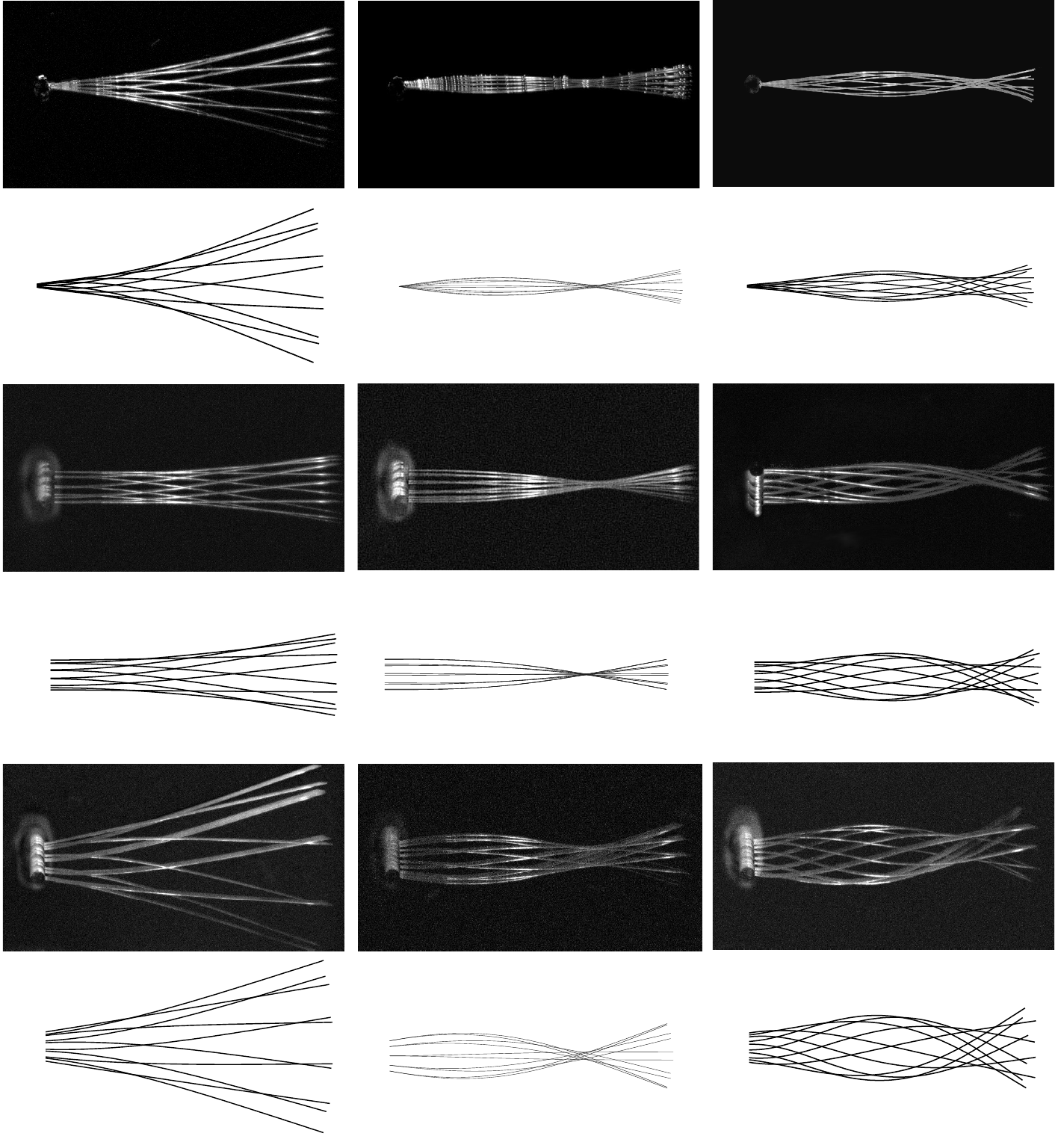}
\caption{Superimposed snapshots of the oscillating plate in the experiment showing the envelopes of three different deformation modes (first, third and fifth rows), compared to the corresponding numerical simulations (second, fourth and sixth rows). Left and right columns correspond to the first and second resonant modes at $\omega^*=0.2$ and 2, respectively. The center column at $\omega^*=1$ is an out-of-resonance case.  First and second rows: pitching; third and fourth rows: heaving; fifth and sixth rows: pitching and heaving at $\phi=0$.  }
\label{Fig_kinematics_experiment}
\end{figure}

\section{Experimental model}

In parallel, experiments with a rectangular elastic plate actuated at one of its extremities were performed on the platform described in \cite{Raspa:2014}, slightly modified in order to study pitching and heaving actuation. The foil is held at constant depth in a water tank by means of an air-bearing rail (see Figure \ref{Fig_setup}), which restricts motion along one direction. Thrust force was measured using a load cell (FUTEK LSB200 Miniature S-Beam with a 50 Hz low-pass filter in the signal conditioner) to hold the cart of the air-bearing rail at a fixed station. High-speed video recordings at 100Hz were used to monitor the deformation kinematics. The elastic plate was made of Mylar of thickness $h=175\mu$m and bending rigidity $B=2.6e^{-3}$N m. The length of the plate was of $L=0.12$m and the span $H=0.04$m, giving an aspect ratio $H^*=0.3$. The amplitudes of the forcing were chosen at a value small enough to allow for a reasonable comparison with the weakly nonlinear model of section \ref{sec:model}: The pitching angle was fixed at  $\psi_0=10^{\circ}$ while the heaving displacement was of $A_0=6$mm, corresponding to a dimensionless heaving amplitude $A_0^*=0.05$. The three different actuation modes described by equations \ref{border_cond} were tested: heaving, pitching and the combination of both, the latter case with the two components of the motion set in phase ($\phi=0$). 

The frequency of the forcing was varied from $f=$0.3 to 3 Hz, which allowed us to observe the transition from the first to the second modes of deformation (see Figure \ref{Fig_kinematics_experiment}). The Reynolds number based on the lateral movement of the plate is here in the range $10000\lesssim Re_w\lesssim 90000$.

\section{Results}\label{sec:results}
\subsection{Plate kinematics and average thrust: model versus experiment}

\begin{figure}
\centering
\begin{tabular}{c c c}
(a) & (b) & (c)\\
\includegraphics[width=0.31\textwidth]{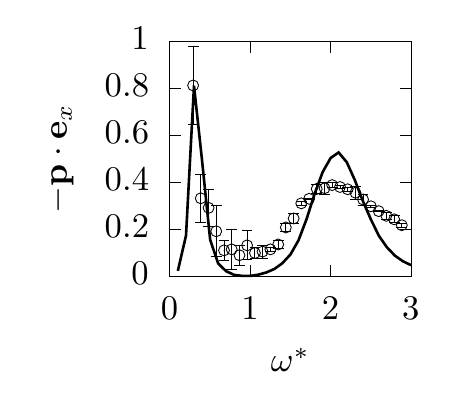}&
\includegraphics[width=0.31\textwidth]{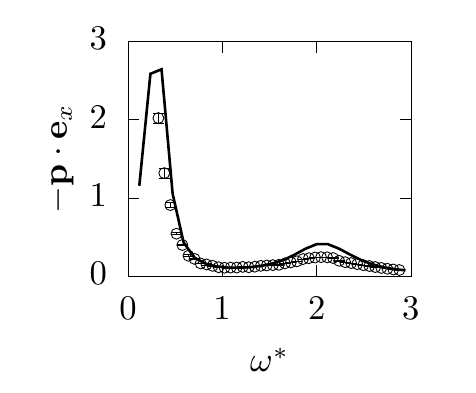}&
\includegraphics[width=0.31\textwidth]{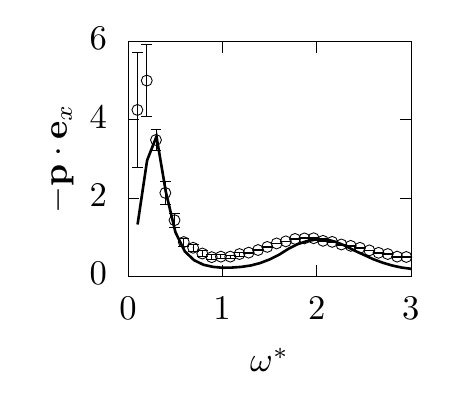}\\
\end{tabular}
\caption{Thrust production of  (a) pure heave, (b) pure pitch and (c) pitch and heave motions.  Comparison between the experimental force measurements (symbols) and the output of the model (lines).}
\label{Fig_thrust_exp_vs_simu}
\end{figure}

Figure \ref{Fig_kinematics_experiment} presents a qualitative comparison between the experiments and the theoretical model showing the kinematics of the plate for two representative cases (mode 1 and mode 2) of the three different actuations. It can be seen that in all cases the simulation reproduces well the main features of the observed deformations. More quantitatively, Figure \ref{Fig_thrust_exp_vs_simu} shows the average thrust force as a function of the dimensionless frequency $\omega^*$ for both the output of the force sensor in the experiments and the numerical solution of the theoretical model. The resonant behaviour of the system is clearly visible in these plots \cite[see also][]{Michelin:2009,Leftwich:2012,Dewey:2013,Paraz:2016}, where local peaks in thrust appear for forcing frequencies corresponding to the different bending modes.  

Based on the good agreement between experiments and numerical simulations, we will further use our model to study in more detail the thrust production mechanisms. We will focus on a self-propelled regime and examine the different types of actuation (heave, pitch, and a combination of both).

\subsection{Thrust production and self-propelled velocity}

The numerical model was used to compute the self-propelled cruising speed that the different actuations would produce, using the skin friction model of \cite{Ehrenstein:2014} to complete the force balance in the swimming direction.  

\begin{figure}
\begin{tabular}{c c}
(a)&(b)\\
\includegraphics[width=0.45\linewidth]{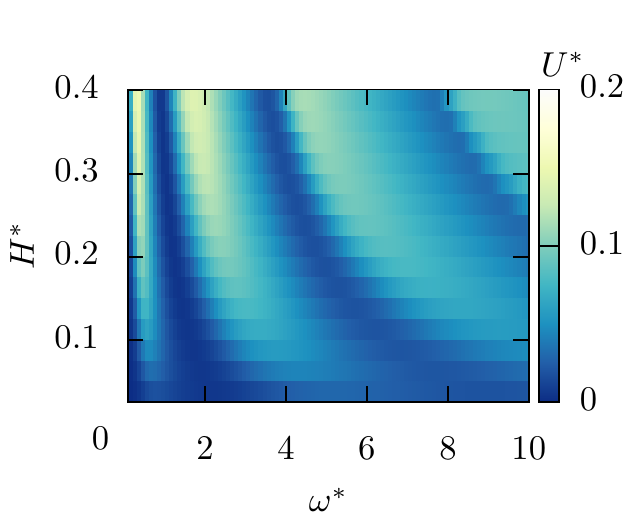}&
\includegraphics[width=0.45\linewidth]{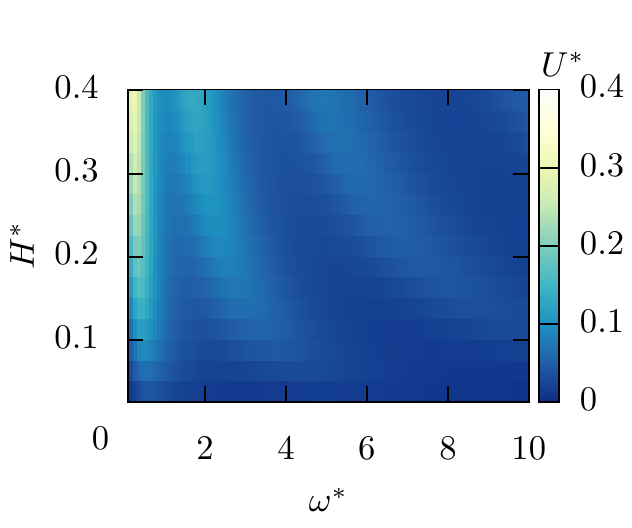}\\
(c)&(d)\\
\includegraphics[width=0.45\linewidth]{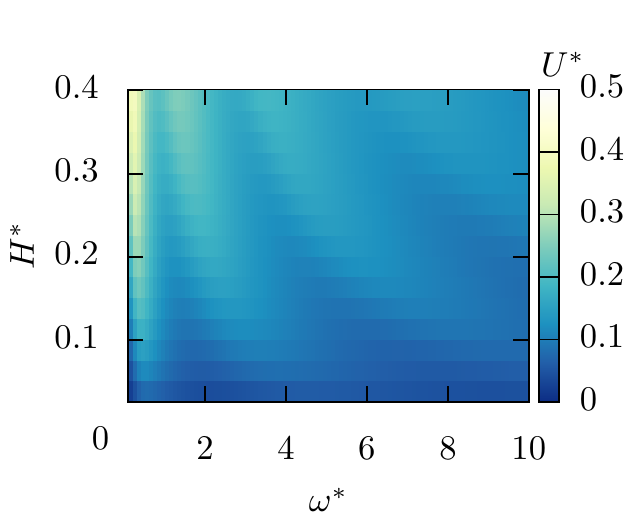}&
\includegraphics[width=0.45\linewidth]{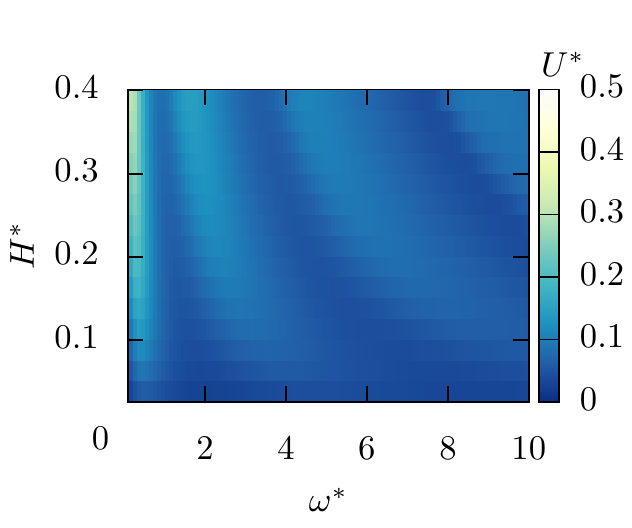}\\
\end{tabular}
\caption{Cruising speed $U^*$ predicted by the numerical model as a function of the non-dimensional flapping frequency $\omega^*$ and the plate aspect ratio $H^*$ for (a) pure heave, (b) pure pitch, (c) pitch and heave with zero phase lag, i.e. $\phi=0$, and (d) pitch and heave with $\phi=-\pi/2$.}\label{fig:U}
\end{figure}

The cruising speed as a function of the aspect ratio of the plate and the non-dimensional frequency $\omega^*$ is presented in figure \ref{fig:U} for four different actuations: (a) heave, (b) pitch, (c) heave and pitch at $\phi=0$, and (d) heave and pitch at $\phi=-\pi/2$. Of course, these maps give a clear picture of the total thrust produced, which is balanced by the skin friction.  Again, the resonant bands appear clearly, larger amplitudes of deformation producing higher thrust and leading to higher cruising speed in all cases. The difference between the heave only and pitch only cases shows that the outcome in terms of performance is actually very dependent on the kinematics, the pitch only case being able to produce significantly larger cruising speeds than the heave only case. Not surprisingly, the case actuated with heave and pitch at $\phi=0$ produces the largest swimming speeds, as the combination of both motions offers a larger swept amplitude compared to the two other actuations. The Reynolds number range based on the displacement speed is here in the range $2000\lesssim Re_l \lesssim 30000$

\subsection{Reactive and resistive thrust}

\begin{figure}
\begin{tabular}{c c}
(a)&(b)\\\includegraphics[width=0.45\linewidth]{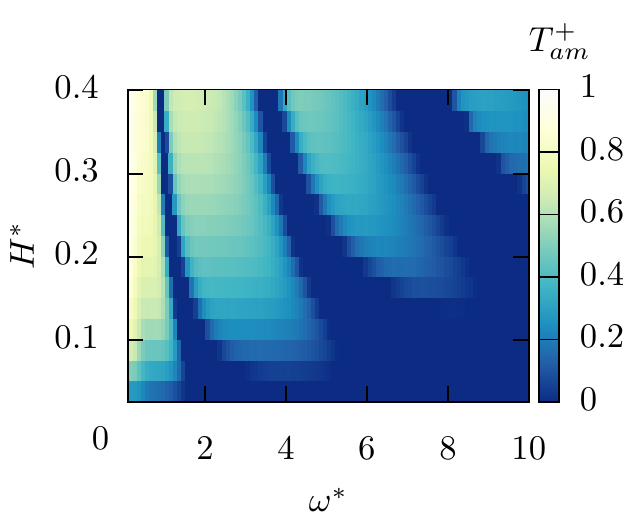}&
\includegraphics[width=0.45\linewidth]{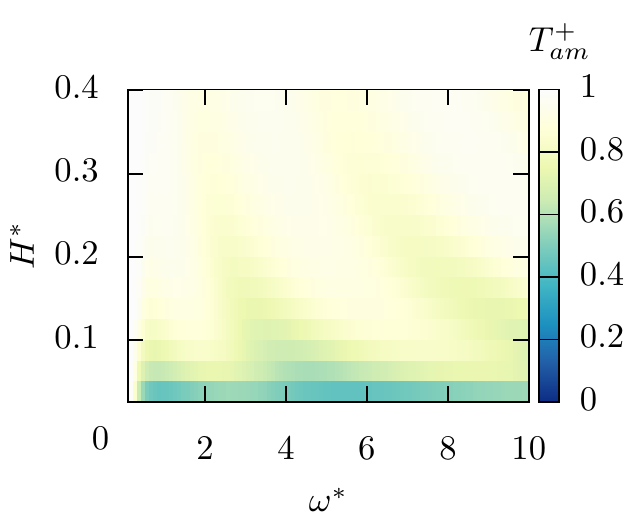}\\
(c)&(d)\\
\includegraphics[width=0.45\linewidth]{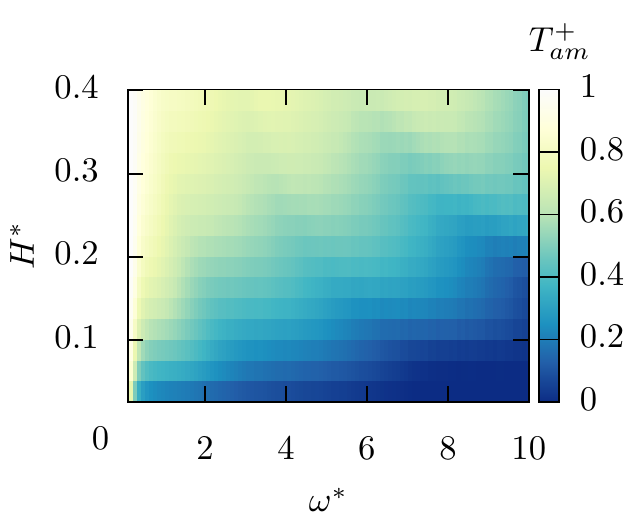}&
\includegraphics[width=0.45\linewidth]{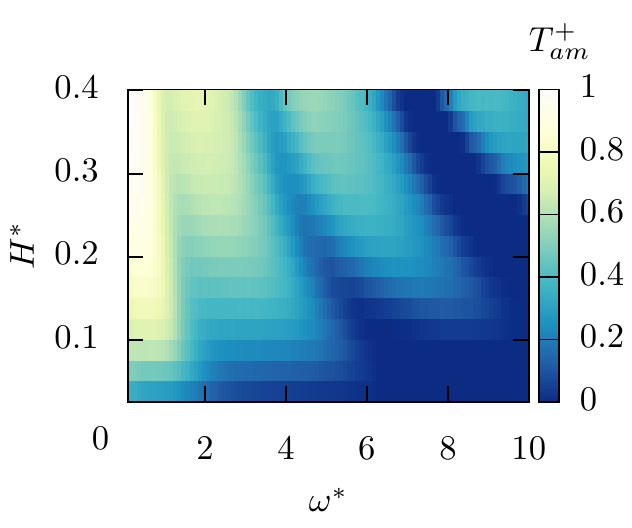}\\
\end{tabular}
\caption{Ratio of reactive thrust to total thrust ${T}_{am}^+$ for the simulations in self-propelled configuration as a function of the non-dimensional flapping frequency $\omega^*$ and the plate aspect ratio $H^*$: (a) pure heave, (b) pure pitch, (c) pitch and heave with zero phase lag, i.e. $\phi=0$, and (d) pitch and heave with $\phi=-\pi/2$. }\label{fig:T}
\end{figure}

As said before, the cruising velocity in figure \ref{fig:U} is the outcome of thrust production mechanisms of two distinct physical origins: the resistive and the reactive terms defined in equations \ref{P_d} and \ref{P_am}. Figure \ref{fig:T} shows maps in the dimensionless frequency versus aspect ratio parameter space of the ratio 

\begin{equation}
{T}_{am}^+=\left<\frac{\left[\int_0^Lp_{am}ds\right]^+}{\left[\int_0^Lp_{am}ds\right]^++\left[\int_0^Lp_{d}ds\right]^+} \right>
\label{Tamplus}
\end{equation}

\noindent which is the normalised temporal mean of the global reactive force production in the direction of motion, the $\left[\cdot\right]^+$ symbol meaning that only the positive contributions to the integral of $-\mathbf{p}_{am}\cdot\mathbf{e}_x$ and $-\mathbf{p}_{d}\cdot\mathbf{e}_x$  are considered. Equation \ref{Tamplus} represents thus the percentage of the total thrust produced by the added mass mechanisms and it is slightly different from the definition

\begin{equation}
{T}_{am}=\left<\frac{\int_0^Lp_{am}ds}{\int_0^Lp_{am}ds+\int_0^Lp_{d}ds} \right>
\label{Tam}
\end{equation}

\noindent used by \cite{Eloy:2013} and \cite{Pineirua:2015}, which can produce values larger than 1 when the resistive component is producing drag and not thrust, and negative values when the reactive thrust is negative while the total thrust is positive. We have chosen to represent $T_{am}^+$ in figure \ref{fig:T} to give a comparative view of the four cases focusing on a measurement of the reactive thrust production that ranges between 0 and 1 in all cases, but we will also use $T_{am}$ in the discussion. It can be readily seen that the balance between the reactive and resistive contributions to the total thrust strongly depends on the geometry and kinematics of the swimming motion, represented here by the aspect ratio and the forcing frequency, respectively.  In the pure heaving case (figure \ref{fig:T}-a), there is a manifest modal signature, the resonant regions of the parameter space characterised by a larger influence of the reactive thrust production. It should be noted that out of resonance, the swimmer rapidly gets into a regime where ${T}_{am}\rightarrow 0$ and the resistive thrust production dominates.  This ``out of resonance'' regime is however highly inefficient in the case of purely heaving motion, as can be seen by the evident reduction of the cruising speed (figure \ref{fig:U}-a). On the other hand, resistive thrust is actually dominant at all frequencies for the lowest aspect ratios ($H^*\lesssim0.1$). The picture for the pure pitching case is different (figure \ref{fig:T}-b), with the reactive contribution to the thrust dominating everywhere, except for the very thin swimmers ($H^*\lesssim0.05$). In the cases with combined pitch and heave actuation the role of the phase lag between the heave and pitch components $\phi$ is crucial. At $\phi=0$ (figure \ref{fig:T}-c) we encounter a more hybrid map, with no modal signature, where larger aspect ratios dominated by the reactive term while lower aspect ratios are dominated by the resistive term. At $\phi=-\pi/2$ (figure \ref{fig:T}-d) the modal signature is also almost erased for the first mode but reappears for the second and third modes.

\section{Discussion}\label{sec:discussion}

We can first remark that, as expected, the reactive contribution decreases for low aspect ratios of the plate ($H^*\leq0.15$) where resistive effects start to dominate thrust. This result has been well known for decades \cite[][]{Lighthill:1960}, since the added mass contribution  scales linearly with span (see Eq. \ref{P_am_adim}).  However, the results presented in figure \ref{fig:T}  show  that the different actuations  of the elastic swimmer have a significant effect on the balance between resistive and reactive force production.  In addition, we can see that for each actuation the transitions between the different deformation modes of the elastic plate can also influence the balance between resistive and reactive contributions to thrust.  In the case of pure pitching the reactive thrust production is always dominant, but the resistive contribution is nonetheless non-negligible for very thin swimmers.  Now, the role of resonances in flexible flapping structures has been well studied recently ---see e.g. \cite{Alben:2008} for pitching actuation and \cite{Michelin:2009,Quinn:2014,Yeh:2014,Yeh:2016,Cros:2016} for heaving actuation--- and the modulation of the thrust (and consequently of the cruising velocity) as a function of frequency observed here, depicting the modal response of the structure, is in agreement with previous observations. However, the strong modal signature in the balance between the reactive and resistive contributions to the thrust in the pure heaving case (figure \ref{fig:T}-a) is a novel observation. The modulation is much less marked in the pure pitching case (figure \ref{fig:T}-b) and it also  depends on the phase lag between the pitch and heave components in the combined actuation (figures \ref{fig:T}-c and \ref{fig:T}-d). In the following we further discuss these results for each type of actuation independently by analysing the role of the trailing-edge amplitude in the production of thrust.

\subsection{Heaving}

\begin{figure}
\centering
\hspace{-1cm}\includegraphics[width=0.95\linewidth]{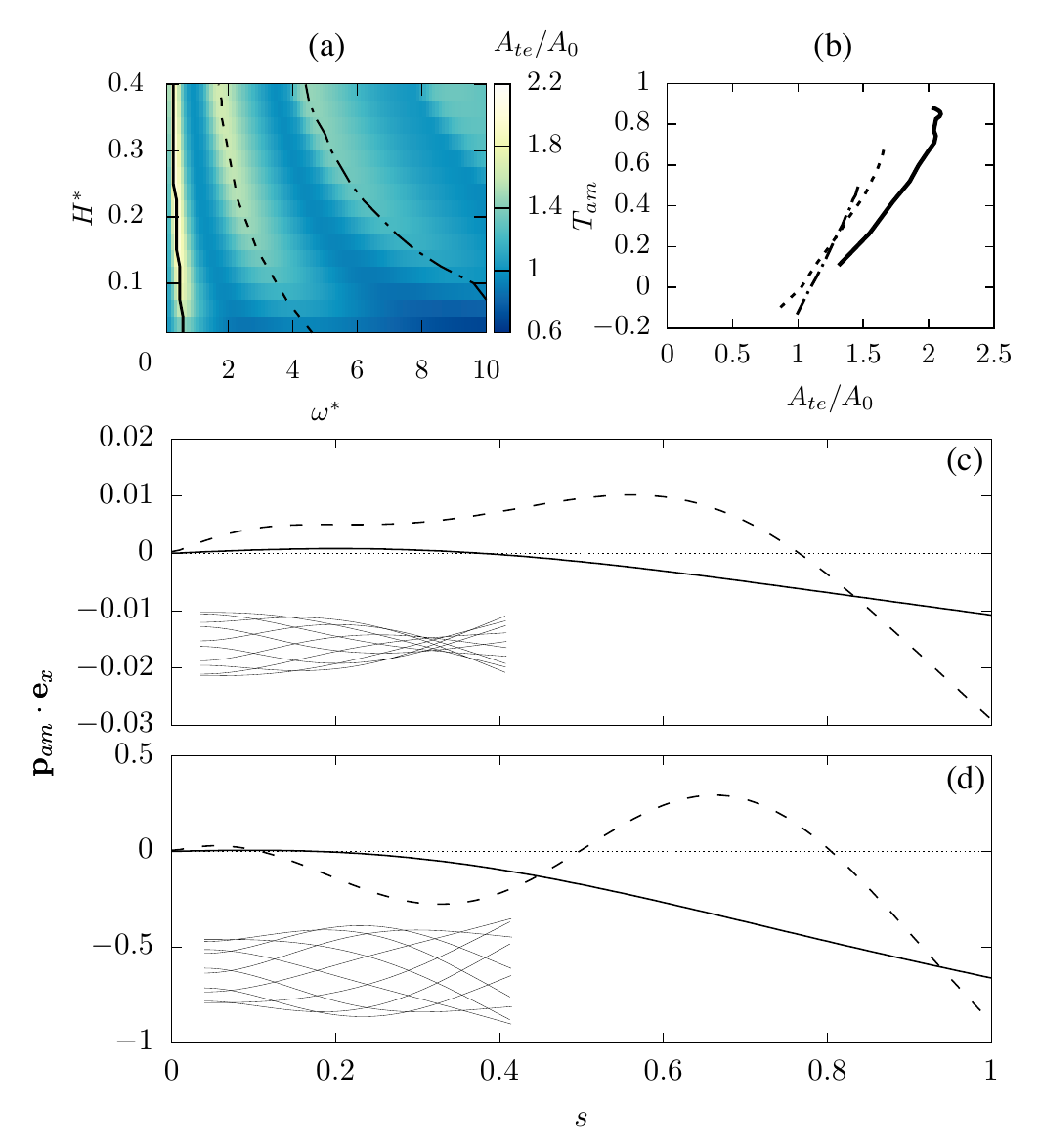}
\caption{Heave case analysis: (a) map of the trailing edge amplitude normalized by the heaving forcing amplitude $A_{te}/A_0$ as a function of the non-dimensional flapping frequency $\omega^*$ and the plate aspect ratio $H^*$; (b) Ratio of reactive thrust to total thrust ${T}_{am}$ as a function of $A_{te}/A_0$ for the first, second and third resonances, respectively plotted in solid, dashed and dashed-dot lines, shown also in (a); (c) and (d) reactive force $\mathbf{p}_{am}\cdot\mathbf{e}_x$ as a function of the curvilinear coordinate $s$ for the two extreme aspect ratios tested: $H^*=0.025$ (c) and 0.4 (d). The first and second modes are shown (in solid and dashed lines, respectively). Snapshots of the deformation dynamics for the second modes are shown as insets. }
 \label{fig:heave_analysis}
\end{figure}

Figure \ref{fig:heave_analysis}-a shows a map of the trailing-edge amplitude $A_{te}$, normalised by the amplitude of the actuation $A_0$. The correlation to the cruising velocity map of figure \ref{fig:U}-a is evident, the larger trailing-edge amplitudes of the resonant modes corresponding to higher velocities. We follow the resonant branches in the map for the first three modes (solid, dashed and dash-dot lines in figure \ref{fig:heave_analysis}-a), and show in figure  \ref{fig:heave_analysis}-b the corresponding reactive contribution to the thrust $T_{am}$. We can see that for the second and third resonant modes the reactive contribution can become negative and, remarkably, this occurs when  $A_{te}/A_0\lessapprox 1$. This can be explained by recalling that the reactive force is the result of integrating $\mathbf{p}_{am}\cdot\mathbf{e}_x$ along the length of the swimming foil.  We will give a detailed explanation of this point further in this section. Figures \ref{fig:heave_analysis}-c and \ref{fig:heave_analysis}-d show  the value of the reactive force production as a function of the curvilinear coordinate $s$ for the two extreme aspect ratios tested (figure \ref{fig:heave_analysis}-c is the case of the thinnest plate, $H^*=0.025$, and figure \ref{fig:heave_analysis}-d that of the widest case, $H^*=0.4$). The first and second resonant modes are shown (solid and dashed lines, respectively). While for the first mode there is thrust production all along the length of the plate, for the second mode we observe that large portions of the plate have on average positive values of $\mathbf{p}_{am}$ (i.e. they produce drag). The deformation dynamics for the second mode is shown in the insets of figures \ref{fig:heave_analysis}-c and \ref{fig:heave_analysis}-d, showing the effect of aspect ratio on the ratio of the trailing-edge to leading-edge amplitudes.  It must be noticed that the difference in the deformation dynamics for plates of different aspect ratios is crucial in the balance between added mass and resistive contributions to thrust.

In his development of the elongated-body theory, \cite{Lighthill:1970} pointed  out that total thrust can be estimated only by considering the motion of the trailing edge.   This conclusion is based on the fact that the added mass coefficient vanishes at the leading edge (i.e. the span $H(s)$ at $s=0$ is zero, as is the case for a fish profile) and thus the leading edge does not contribute to the momentum exchange between the swimming body and the surrounding fluid \cite[see also][]{Eloy:2013}.  In the case of a rectangular plate of constant span, like the one considered in the present study, the contribution of the leading edge cannot be neglected directly. In the small deformation  approximation ($y'\ll1$), taking $U_n\sim\partial y /\partial t + U^*\partial y /\partial s$ and $U_t\sim -U^*$, the integral from $0$ to $L$ of the $x$-projection of (\ref{P_am}) gives :

\begin{equation}
F_p=\frac{1}{2}\mathcal{M}(H)\left(\left[\left(\frac{\partial Y}{\partial T}\right)^2-U^{2}\left(\frac{\partial Y}{\partial S}\right)^2  \right]_0^L - \int_0^L\frac{\partial}{\partial t}\left(\frac{\partial Y}{\partial T}+U^*\frac{\partial Y}{\partial S}\right)\frac{\partial Y}{\partial S}dS\right).
\label{T_light}
\end{equation}
As shown by \cite{Lighthill:1970}, the last term in the previous equation does not contribute to the mean thrust since it is the time derivative of a quantity fluctuating between fixed limits.  Thus the mean propulsive force can be expressed as :  

\begin{equation}
F_p=\frac{1}{2}\mathcal{M}(H)\left<\left[\left(\frac{\partial Y}{\partial T}\right)^2-U^{2}\left(\frac{\partial Y}{\partial S}\right)^2  \right]_0^L\right>.
\end{equation}  

Considering that  in $S=L$,  $\partial Y/\partial T$ scales as $\omega A_{te}$ and $\partial Y/\partial S$ as $A_0/L$, and in $S=0$,  $\partial Y/\partial T$ scales as $\omega A_0$ and, in the case of pure heaving, $\partial Y/\partial S=0$, in the limit $U\ll \omega L$ the total propulsive force then scales as :

\begin{equation}
F_p\sim\frac{1}{2}\mathcal{M}(H) \omega^2 A_0^2\left(\left(\frac{A_{te}}{A_0}\right)^2-1\right),
\end{equation}

\noindent showing that the force $F_p$ contributes to propulsion only if $A_{te}>A_0$, as observed in figure \ref{fig:heave_analysis}-b.   It must then be noticed that in the case of heaving flexible panels, taking into account only the trailing-edge dynamics to estimate the total thrust can lead to errors, and attention must be payed to the global dynamics of the plate.   Moreover, particular attention must be payed to regimes away from the plate resonances, where the trailing-edge amplitudes decrease substantially, leading to added mass drag effects \cite[as noticed  for example by][]{Paraz:2016}.  This effect can also be seen for small aspect ratios  (as shown in figure \ref{fig:heave_analysis}-c), where the dominant dissipation due to the lateral displacements of the plate  tends to reduce the trailing-edge amplitude (as shown in the inset of figure \ref{fig:heave_analysis}-c).

\subsection{Pitching}
\begin{figure}
\centering
\hspace{-1cm}\includegraphics[width=0.95\linewidth]{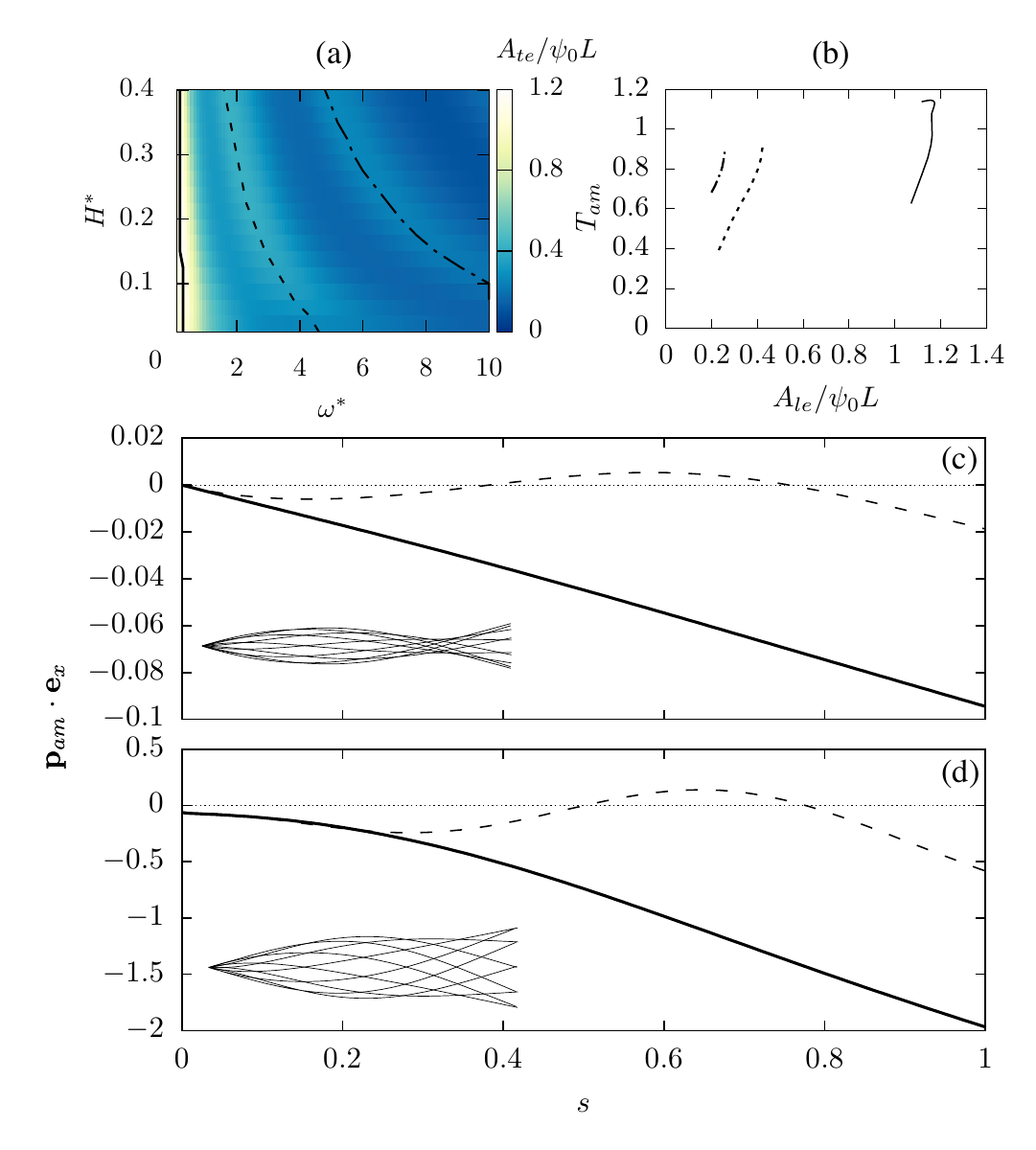}
\caption{Pitch case analysis: (a) map of the normalized trailing-edge amplitude  $A_{te}/\psi_0 L$ (see text) as a function of the non-dimensional flapping frequency $\omega^*$ and the plate aspect ratio $H^*$; (b) Ratio of reactive thrust to total thrust ${T}_{am}$ as a function of $A_{te}/\psi_0 L$ for the first, second and third resonances, respectively plotted in solid, dashed and dashed-dot lines, shown also in (a); (c) and (d) reactive force $\mathbf{p}_{am}\cdot\mathbf{e}_x$ as a function of the curvilinear coordinate $s$ for the two extreme aspect ratios tested: $H^*=0.025$ (c) and 0.4 (d). The first and second modes are shown (in solid and dashed lines, respectively). Snapshots of the deformation dynamics for the second modes are shown as insets. }
 \label{fig:pitch_analysis}
\end{figure}

Figure \ref{fig:pitch_analysis}-a shows a map of the trailing-edge amplitude $A_{te}$, normalised by the hypothetic beating amplitude of an infinitely rigid plate $\psi_0L$. As in the case of heaving, there is a fair correlation to the cruising velocity map (figure \ref{fig:U}-b), the larger trailing-edge amplitudes of the resonant modes corresponding to higher velocities.  For a pure pitching plate we observe a drastic decrease in the trailing-edge amplitude when the plate switches from the first resonant mode to higher modes. This is clear in the $T_{am}$ versus $A_{te}/\psi_0L$ plot of figure \ref{fig:pitch_analysis}-b where the curves for the second and third modes are placed at lower values of  $A_{te}/\psi_0L$. We can thus note that the $\psi_0L$ length scale is a good approximation to the trailing-edge oscillation amplitudes for the first mode (giving values of $A_{te}/\psi_0L$ close to 1), while it significantly overestimates the observed values for higher modes. Two more observations can be made from figure \ref{fig:pitch_analysis}-b: on the one hand, the first resonant mode $T_{am}$ is sometimes larger than 1 (the region where  $T_{am}>1$ corresponds to high aspect ratios), which means, as mentioned before, that the resistive term $\mathbf{p}_d$ contributes to drag; on the other hand, in contrast with the heaving plate, the added mass contribution to thrust is less sensitive to plate deformation mode transitions and is, in general, higher than that for a heaving plate.  This can also be seen in the added mass force distribution as a function of $s$ shown in figures \ref{fig:pitch_analysis}-c and \ref{fig:pitch_analysis}-d .  In both graphs, corresponding to the two extreme aspect ratios, the distribution of $\mathbf{p}_{am}$ along $s$ has a similar shape, the main difference being its value, which is of course higher for the larger plate.  

\begin{figure}
\centering
\hspace{-1cm}\includegraphics[width=0.85\linewidth]{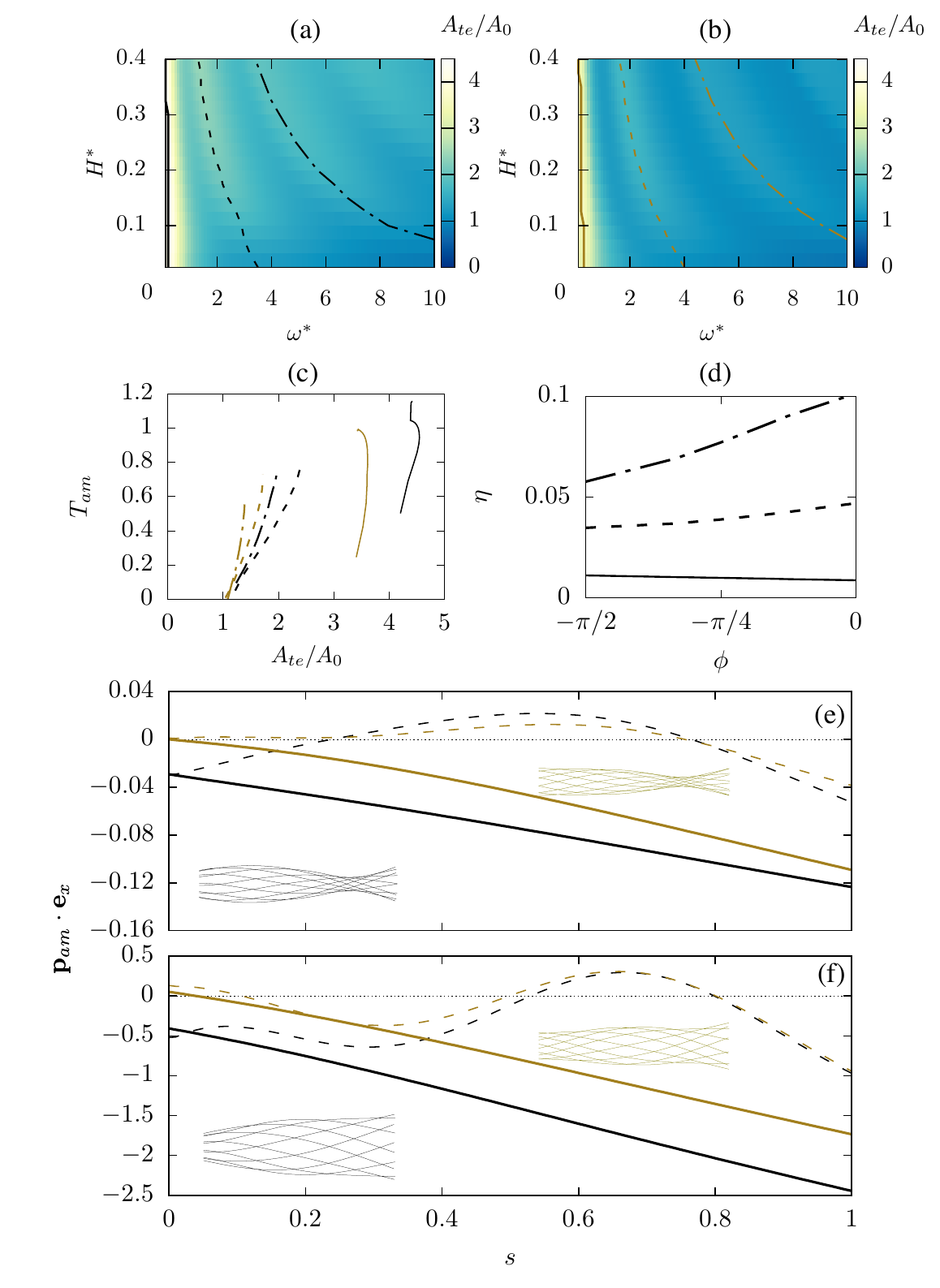}
\caption{Pitch-heave analysis: (a) and (b) maps of the normalized trailing-edge amplitude  $A_{te}/A_0$ as a function of the non-dimensional flapping frequency $\omega^*$ and the plate aspect ratio $H^*$ for (a) $\phi=0$ and (b) $\phi=-\pi/2$; (c) Ratio of reactive thrust to total thrust ${T}_{am}$ as a function of $A_{te}/A_0$ for the first, second and third resonances, respectively plotted in solid, dashed and dashed-dot lines;  (d) propulsion efficiency as a function of phase lag; (e) and (f) reactive force $\mathbf{p}_{am}\cdot\mathbf{e}_x$ as a function of the curvilinear coordinate $s$ for the two extreme aspect ratios tested: $H^*=0.025$ (e) and 0.4 (f). The first and second modes are shown (in solid and dashed lines, respectively). Snapshots of the deformation dynamics for the second modes are shown as insets. Black lines: $\phi=0$; Grey lines: $\phi=-\pi/2$. }
 \label{fig:pitch_heave_analysis}
\end{figure}

\subsection{Pitching and heaving}

The equivalent analysis for the cases with combined pitching and heaving actuation is presented in figure \ref{fig:pitch_heave_analysis}. Of course, several points that have been discussed for the only heaving and only pitching cases appear also here. In both cases we find again in the maps of the trailing-edge amplitude $A_{te}$ (figures \ref{fig:pitch_heave_analysis}-a and \ref{fig:pitch_heave_analysis}-b) the correlation with the cruising velocity maps (figures \ref{fig:U}-c and \ref{fig:U}-d, respectively). We may note that the case with $\phi=0$ reaches higher velocities than that of $\phi=-\pi/2$. We can also observe the decrease in the trailing-edge amplitude when the plate switches from the first resonant mode to higher modes (figure \ref{fig:pitch_heave_analysis}-c, inherited from the pitching component of the motion). But the crucial point concerning the combined pitching and heaving actuation is the role of the phase lag $\phi$ between the two components of actuation. We have chosen to examine closely two cases: $\phi=0$ and $\phi=-\pi/2$. Already in the maps of $T_{am}$ in figures \ref{fig:T}-c and \ref{fig:T}-d it can be seen that the contribution of reactive thrust changes remarkably with the phase lag: at  $\phi=0$, the map has almost lost all modal signature, while at $\phi=-\pi/2$ the resonant effect appears clearly. The origin of this difference can be elucidated looking at figures \ref{fig:pitch_heave_analysis}-e and \ref{fig:pitch_heave_analysis}-f. For the two extreme aspect ratios depicted, the effect of the phase lag in the first mode is qualitatively the same, the case of $\phi=-\pi/2$ shifting the absolute value of $\mathbf{p}_{am}\cdot\mathbf{e}_x$ to lower values with respect to the case of $\phi=0$ (bringing it close to zero near the leading edge of the plate). For the second mode the picture is different: we observe that added mass contributions to thrust in the rear half-part of the plate (towards the trailing edge) are almost the same despite the phase lag.  However, in the frontal part of the plate there is an important difference, mainly near the leading edge.  Whilst for the case with a phase lag $\phi=-\pi/2$ the first section of the plate does not contribute to the thrust, even generating drag slightly,  for the case with no phase lag the frontal part of the plate contributes substantially to the thrust. This may be explained by the synchronisation between maximum acceleration and maximum angle at the leading edge, which is higher in the $\phi=0$ case.

The phase lag has also relevant implications in the propulsion efficiency of flexible actuated plates.  For a free swimming flexible plate, following the definition given by \cite{Zhang:2010}, the efficiency can be estimated as the ratio between the kinetic energy $E=1/2\mu U^2$ (for a plate of mass per unit surface $\mu$ travelling at speed $U$) and the work per cycle $W$ done by the actuation at the leading edge :
\begin{equation}
 W=\int_T^{T+\Delta T}\frac{\partial Y}{\partial T}(0)\left(F_{\tau}(0)\boldsymbol{\tau}-B\frac{\partial ^2\theta}{\partial S^2}(0)\mathbf{n}\right)\cdot \mathbf{e}_y dT+\int_T^{T+\Delta T}\frac{\partial \theta}{\partial T}(0)B\frac{\partial \theta}{\partial S}(0)dT.
\end{equation}

In terms of non-dimensional  parameters, the efficiency can be written as :

\begin{equation}
\eta=\frac{U^{*2}}{2w}
\end{equation}

\noindent where 

\begin{equation}
w=\int_t^{t+2\pi}\frac{\partial y}{\partial t}(0)\left(f_{\tau}(0)\boldsymbol{\tau}-\frac{1}{\omega^{*2}}\frac{\partial ^2\theta}{\partial s^2}(0)\mathbf{n}\right)\cdot \mathbf{e}_y dt+\int_t^{t+2\pi t}\frac{\partial \theta}{\partial t}(0)\frac{1}{\omega^{*2}}\frac{\partial \theta}{\partial s} (0) dt.
\end{equation}

In figure \ref{fig:pitch_heave_analysis}-d we present the evolution of the efficiency as a function of $\phi$.  Except from the first resonant mode, which presents a slight increase in efficiency as the phase lag goes from $0$ to $-\pi/2$, efficiency tends to be higher when there is no phase lag between the heaving and pitching actions. This can be a surprising result if one thinks of rigid foils, where $-\pi/2$ or values close to that have been many times shown to be optimal in terms of efficiency, or at most  in flexible wings only deforming in the first mode such as insect wings \cite[see e.g.][]{Kang:2013}. And indeed, in a recent study with flexible plates with combined actuation \cite{Quinn:2015} found an effect of phase lag opposite to what we observe here.  In fact, for $\phi=-\pi/2$ the angle of attack is minimised, and as a consequence, the work done by the actuator at the leading edge is smaller.   However, the $-\pi/2$ phase lag reduces the trailing-edge amplitude, and thus, the generated thrust tends to be smaller.  In the study of  \cite{Quinn:2015} this competition between the effects of angle of attack and trailing-edge amplitude is evidenced by a reduction of the optimal phase, which in some cases happens to be slightly smaller than $\pi/2$.  The latter is not observed in our simulations.  For the second and third modes we clearly observe that efficiency increases as the phase lag tends to zero.  This could mean that, for the parameters used in our study, the effects of amplitude reduction dominate with respect to those relative to the angle of attack.  One main difference between our study and previous works is the aspect ratio of the plate.  In all previous works confirming the maximum efficiency for $\phi=-\pi/2$, the plate aspect ratio is $H^*>0.75$.  To our knowledge, there is no experimental evidence that for aspect ratios $H^*<0.4$ the phase lag--efficiency relation holds true.  Unfortunately, the experimental verification of the latter is out of scope of the present work.  Some other differences, such as the small heaving and pitching actuations used in our study, could also explain the discrepancy of our results.  \cite{Quinn:2015} show that the gain in efficiency with phase lag is lower for smaller pitching angles.  The latter, along with the possible impact of the aspect ratio,  should indeed  be explored experimentally.

\begin{figure}
\centering
\begin{tabular}[t]{c c}
\hspace{0.75cm}(a) &\hspace{0.75cm} (b) \\
\includegraphics[width=0.48\linewidth]{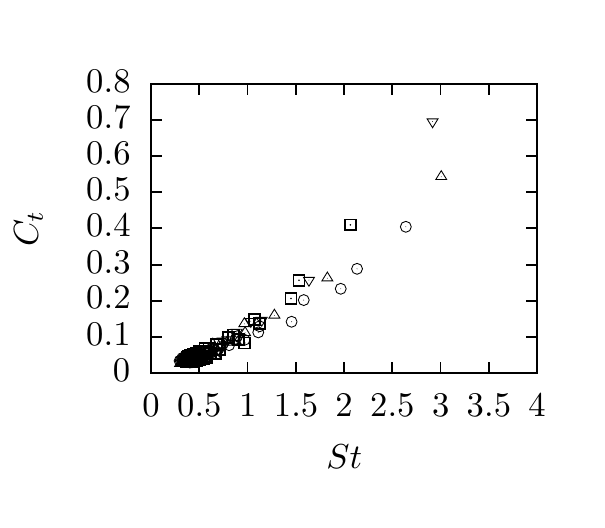}&
\includegraphics[width=0.48\linewidth]{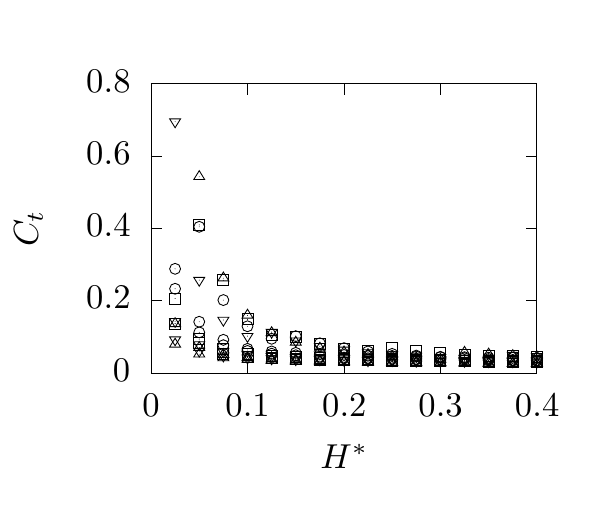}
\end{tabular}
\caption{Thrust coefficient $C_t$ as a function of (a) the Strouhal number $St$ and (b) aspect ratio $H^*$. Different symbols correspond to different actuations: heaving ($\boxdot$), pitching ($\odot$) and combined heaving and pitching at $\phi=0$ ($\triangle$) and $\phi=-\pi/2$ ($\bigtriangledown$)}
 \label{fig:Ct_strouhal}
\end{figure}

\subsection{Thrust coefficient versus Strouhal number}

As last part of our analysis, we examine the evolution of the thrust coefficient 

\begin{equation}
C_t=\frac{(\mathbf{P}_{am}+\mathbf{P}_{d})\cdot\mathbf{e}_x}{\frac 1 2 \rho U^2} \;,
\label{Ct}
\end{equation}
as a function of the the different types of actuation and plate parameters. Figure \ref{fig:Ct_strouhal}-a shows the values of $C_t$ as a function of the Strouhal number $St=2fA_{te}/U=A_{te}/\pi U^* L$ for all different actuations. Only the points corresponding to the resonant curves of the first three modes are plotted. We observe that, in spite of the different contributions from reactive and resistive forces for the different actuations, all cases follow the same scaling in a thrust coefficient versus Strouhal number curve (see Figure \ref{fig:Ct_strouhal}). The data collapse on a  $C_t(St)$ curve is thus more general than what has been attributed in the literature to a purely reactive thrust production mechanism \cite[][]{Quinn:2014}. Remarkably, the changes in the part of reactive and resistive contributions to the total thrust that we have pointed out in the previous sections do not influence the dependence of $C_t$ on $St$.  Now, figure \ref{fig:Ct_strouhal}-b shows $C_t$ as a function of the aspect ratio $H^*$. The plot shows that $C_t$ diminishes with increasing aspect ratio (because $U$ is changing with $H^*$). Larger aspect ratio foils also show less scatter in $C_t$ as has been observed by \cite{Raspa:2014}, who also studied the finite-size effect on the drag on a self-propelled flexible swimmer. Because $C_t=C_d$ in the self-propelled configuration, the behaviour of $C_t$ observed in figure \ref{fig:Ct_strouhal}-b shows that this finite-size effect is governed by the term depending on $\sim1/\sqrt{H}$ in Eq.  \ref{skin_friction_ehrenstein} of the skin friction with boundary-layer thinning of \cite{Ehrenstein:2013}.

\section{Concluding remarks}

The flexible plates studied here can be thought of as a simplified model of a full swimmer or of a flapping propulsor. We show that the appropriateness of using a purely reactive model to predict thrust force production is largely dependent on the geometry and kinematics of the swimmer in question. The use of a complete model including resistive and reactive forces such as the one presented here is thus crucial to obtain an accurate description; in particular in the context of designing realistic bio-inspired applications where geometry and kinematics will define the optimisation parameters. We point out the specificity of these model flexible swimmers, where the kinematics is the passive outcome of a strongly coupled fluid--structure interaction problem, which differentiates them from animal swimmers where muscle action distributed all along the body can impose a greater control on the kinematics. 


\end{document}